\documentstyle[preprint,aps]{revtex}

\newcommand{\be}{\begin{equation}}
\newcommand{\ee}{\end{equation}}
\newcommand{\bea}{\begin{eqnarray}}
\newcommand{\eea}{\end{eqnarray}}

\begin{document}
 \draft
\preprint{\small gr-qc/9505024 Alberta-Thy-34-94}

\title
 {On critical behaviour in gravitational collapse}
\author{Viqar Husain\footnote{email: vhusain@math.ucalgary.ca,
martinez@phys.ualberta.ca, nunez@xochitl.nuclecu.unam.mx},
Erik  A. Martinez$^+$, and Dar\'\i o N\'u\~nez$^\dagger$  }
\address
 {$^*$Department of Mathematics and Statistics,\\
University of Calgary, Calgary, Alberta, Canada T2N 4N1.\\
$^+$Theoretical Physics Institute,\\
 University of  Alberta, Edmonton, Alberta, Canada T6G 2J1.\\
$^{\dagger}$Instituto de Ciencias Nucleares, UNAM,\\
Circuito Exterior CU, A.P. 70-543, M\'exico, D. F. 04510,
M\'exico.}

\maketitle

\begin{abstract}
We give an approach to studying the critical behaviour that has
been observed in numerical studies of gravitational collapse.
These studies suggest, among other things, that black holes
initially form with infinitesimal mass. We show generally how
a black hole mass formula can be extracted from a
transcendental equation.

Using our approach, we give an explicit one parameter set of
metrics that are asymptotically flat and describe the collapse
of apriori unspecified but physical matter fields. The
black hole mass formula obtained from this metric exhibits a
mass gap - that is, at the onset of black hole formation, the
mass is finite and non-zero.

\end{abstract}
\bigskip
\vfill
\eject

A fundamental  problem in general relativity is the
investigation of  the gravitational
collapse of matter fields.  The main
motivation for studying this is the cosmic censorship
conjecture, one form of which states that gravitational
collapse always results in a black hole.

Recently  the collapse problem has been studied numerically
and the results are intriguing. For the spherically symmetric
 collapse of a scalar field,  Choptuik \cite{chop} has  found
that when the initial  matter field  is an ingoing pulse, the
collapsing matter forms a black  hole with  mass  given by
$M = K(c-c_*)^\gamma$, where   $K$ is a constant,  $c$ is any
one of the parameters in the initial data
for the matter field,  $c_*$ is the critical value of this
parameter that gives a zero mass black hole, and
$\gamma\sim .37$.  In particular,  no black hole is formed
when $c<c_*$.   An important feature of this result is that
it appears to be independent  of spherical symmetry and the
type of matter fields: the same mass formula for the
black hole has been found for the axisymmetric  collapse of
gravitational waves by Abraham and Evans\cite{ae} and for the
spherically symmetric collapse of radiation by Coleman
and Evans \cite{ce}.

Thus, these results appear to
reflect a universal property of the Einstein equations in
strong field regions.   So far there is no analytical
understanding of this result,  nor is there an explicit
metric that  exhibits this behaviour.

There have  however been a number of attempts at
explanations, most of which involve self-similar solutions
\cite{oshetal,pat}.  However, the particular  solutions
discussed do not  appear to be relevant for the collapse
problem with asymptotically flat boundary conditions, because
they are cosmological, and the collapse results  in infinite
mass black holes; the radius of the apparent horizon
tends to infinity with time.
 Another  (non-self-similar) time dependent
 solution, given by the authors, has the same shortcoming
\cite{us}.  The problem
has been  discussed in two dimensions where an  exponent of
 $.5$  was  reported  \cite{2d}. There have also been
proposals  for using  perturbations of  black holes to
calculate the exponent analytically  for the  super-critical
 case $c>c_*$, for  which  the black hole always forms
\cite{tras,pull,koike,maison}. Recently, further numerical
perspectives on this problem have been given \cite{gar,eard}.

In this Letter   we present an  alternative analytical
approach to  this problem. The first part of the paper
gives a general approach that may be used to extract a
black hole mass formula from a spherically symmetric time
dependent metric. The mass formula arises from the
solutions of a  transcendental equation. The second part
of the paper gives an explicit example of the procedure:
We give a metric that satisfies the dominant energy
condition and describes a realistic collapse. This specific
example suggests that at the onset of
black hole formation, there may be a mass gap. We comment
on how our approach may be used to get a black hole mass
formula without a mass gap.

We start with the general spherically symmetric line element
\be
ds^2 = -f e^{-2\psi}dv^2 + 2e^{-\psi}dvdr + r^2d\Omega^2
\label{metric}
\ee
where $f(r,v)$ and $\psi(r,v)$ are functions of the  radial
coordinate
$r$, $(0\le r \le \infty)$, and the advanced time coordinate
$v$, $(0\le v \le \infty)$, and
$d\Omega^2 $ is the line element of the unit two-sphere.
  The parametrization (\ref{metric}) has  been used to study
 the collapse of null shells \cite{werner}.
The mass function
$m(r,v)$ is defined by
\be f(r,v) = \bigl( 1 - 2m(r,v)/r \bigr),  \label{mfn}
\ee
and is a measure of the mass contained within radius $r$.
The Schwarzschild solution results from setting
$m(r,v) =$ constant and $\psi(r,v)=0$.
 The mass function will in general be parametrized by
initial data parameters $c_i$.  For simplicity we will assume
dependence on only  one parameter, $c$.

The apparent horizon is the three-surface that separates
regions containing trapped surfaces from the normal regions
of the spacetime. This surface is defined by the equation
$g^{ab}\partial_a r \partial_b r =0$.  For the  metric
(\ref{metric}) it is given by
\be
m(r,v;c) = {r\over 2}.
\ee
 We are interested in the asymptotic $v\rightarrow\infty$
solutions of this equation because this gives the large
time behaviour of the apparent horizon, and this is
what has been investigated numerically \cite{chop,ae}.
Taking this  limit results in the one parameter family
 of transcendental  equations
\be
M(r;c) : = \lim_{v \rightarrow \infty} m(r,v;c)={r\over 2}.
\label{trans}
\ee
The solutions of this equation give  the radial coordinate of
the asymptotic $(v\rightarrow\infty)$ apparent horizon as
a function of $c$. In general there will be ranges of $c$
for which there are no solutions to (\ref{trans}). This will
give the subcritical region where no black hole forms.
Similarly there will be critical and supercritical solutions.

In the numerical work cited above, the
black hole mass is defined by the radius of this asymptotic
 apparent  horizon. This is a reasonable definition because
it is in this limit that the apparent horizon approaches the
event horizon \cite{hawkell}.
In  our discussion,  this mass  is given explicitly
 by the solutions  $r_{AH}(c)$ of these  transcendental
equations,  namely
  \be
M_{BH}(c) := {r_{AH}(c) \over 2}. \label{mbh}
\ee
A plot of the solution of the equations (\ref{trans}) as a
function of $c$ will  give $M_{BH}(c)$.
{\it  We emphasize that the steps outlined above
are general in the sense that if one is given an exact
collapse solution, this procedure may be used to see if
critical behaviour exists in the long time limit}. Our
purpose in this paper is to study what specific mass
formulas can be obtained from (\ref{trans}).

The mass function in the metric (\ref{mfn}) must satisfy certain
physical conditions in order to define a realistic collapse.
These are: (i) The  metric should be  asymptotically
flat, (ii) the mass function should satisfy $m\ge 0$ and
$\partial m/\partial r \ge 0$, and
 should give, in the $v\rightarrow \infty$ limit, a set of
transcendental equations from which we can obtain
 $M_{BH}(c)$, and (iii) the mass function
should increase with advanced time corresponding to an implosion
of matter, at least initially.  Also, if the mass function
leads to multiple apparent horizons, the outermost one will
serve to extract a mass formula.

 We are interested  mainly in the supercritical case, where a
black hole always forms at the end point of collapse, since the
main goal of our work is to extract a black hole mass
formula. However, as we will see, the subcritical case
(no black hole) also arises naturally.

The  Einstein equations  $G_{ab}=8\pi T_{ab}$ for the metric
(\ref{metric})  are
\bea
m^\prime  &=& -4\pi r^2 T_v^{\ v}, \label{mpr} \\
\dot{m}   &=&  4\pi r^2 T_v^{\ r}, \label{mdot}\\
\psi^\prime &=& 4\pi r T_{rr}, \label{psipr}
\eea
where the prime and dot denote $r$ and $v$ derivatives.
We will not fix the stress-energy tensor by specifying
any specific type of matter, but  will instead determine
this tensor in terms of $m$ and $\psi$. The goal then
will be to see if these functions can be fixed so that the
energy conditions for realistic matter are satisfied.
(We note that $\dot{\psi}$ is not fixed by the Einstein
equations).

As an example of the above procedure, we now construct one metric
that describes a realistic collapse.
 This requires  a mass function satisfying
the above three physical conditions, with $\psi$ still arbitrary.
A  suitable choice is
\be
m(r,v;c) = c\ [{\rm tanh}( e^c\  \ln r) + 1]\ {\rm tanh} v
\  .
\label{ourm}
\ee
The Arnowit-Deser-Misner (ADM) mass for this choice is  $2c$.

 The one parameter $(c)$ set of transcendental equations
(\ref{trans})
resulting from this in the $v\rightarrow \infty$ limit are
\be
 c\  [{\rm tanh}( e^c\ \ln r) + 1] = {r\over 2},
\ee
which give,  using (\ref{mbh}), the equation for the
black hole mass $M_{BH}$:
\be
 c\ ({\rm tanh}[ e^c\ \ln (2  M_{BH})] + 1 ) = M_{BH}.
\label{mtran}
\ee

Equation (\ref{mtran}) may be solved numerically, and there
are solutions for $c\ge c_* =.465727$. In the supercritical region
$c > c_*$, the solutions are fit approximately by the equation
\be
 M_{BH} = 0.64 + 20.69 (c-c_*) - 2205.15  (c-c_*)^2
+ 113620 (c-c_*)^3
 \  .
 \label{fit}
\ee
A plot of $M_{BH}$ vs. $(c-c_*)$ is given in  Fig. 1. The
solutions of the transcendental equations (\ref{mtran}) and the
fit (\ref{fit}) were found using  Mathematica, and also checked
with Maple. (The Mathematica commands we used for Fig. 1 are:
tt := Table[$\{$ ($c-.465727$),  (M)/.
FindRoot[ $c$ (Tanh[ Exp[c]\, Log[2M]] +1) == M,
$\{$M, 0$\}$] ]$\},\{c$, .465728,.475000, .0001$\}$]
Fit[ tt, $\{ 1,x,x^2,x^3 \},x$ ].)

There are several comments in order for the fit (\ref{fit}):
(i)  The mass formula {\it is not} a pure power law as observed
in the numerical integrations of the scalar field and
null fluid \cite{chop,ce}. (ii) This particular example
exhibits a mass gap, because at criticality Eqn. (\ref{mtran})
gives $M_{BH} = .64$, as Fig. 1 clearly shows.
(iii) For $c < c_* $ there are no solutions of
(\ref{mtran}), and hence no black hole formation. (iv) For
$c > c_*$ there are two intersections of the mass function
with the $r/2$ curve, and hence an inner and outer apparent
horizon. The above fit corresponds to the outer horizon,
which is what is relevant for determining black hole masses.

In summary, although this mass function example does not give
the mass formula  obtained in the full numerical
integrations \cite{chop,ae,ce}, it does give the
subcritical, critical and supercritical regions.
 Furthermore, as we will see below, this
mass function corresponds to matter satisfying the
  dominant energy conditions, and thus gives a
{\it physical} asymptotically flat collapse solution.
This suggests that there are physical solutions that exhibit
a mass gap.

 So far  the metric function $\psi(r,v)$, is arbitrary.
The task now is to see how this function
 is restricted so that the metric arises
  from realistic matter.  We would like to give at
least one example of such a  $\psi(r,v)$ in order to
have an explicit metric. To do this  we first turn to a
discussion of the energy conditions.

The energy conditions may be  imposed most simply by first
diagonalizing the stress-energy tensor, (determined in
terms of  $m$ and $\psi$), by solving
$T_a^{\ b}v_b = \lambda v_a$.  Diagonalizing the $(v,r)$
part of  $T_a ^{\  b}$  gives the eigenvalues
$\lambda_{\pm}$:
\be
4\pi r^2 \lambda_{\pm} =  -m^\prime + {rf\psi^\prime\over 2}
\bigl[\ 1 \pm
\sqrt{ 1 + { 4\dot{m} e^{-\psi} \over r f^2 \psi^\prime} }
\ \bigr]\ .
\label{eval}
\ee
The corresponding   $(v,r)$ components  of the eigenvectors
are
\be
u_a^{(\pm)} = ( 1 , { 4\pi r^2 \lambda_{\pm} +
m^\prime \over \dot{m} } ), \label{evec}
\ee
and their norms are
\be
| u^{(\pm)} | = \bigl( {4\pi r^2 \lambda_{\pm} + m^\prime
\over \dot{m} } \bigr)
\bigl[\ 2 e^{-\psi}  +
 f \bigl( {4\pi r^2 \lambda_{\pm} + m^\prime
\over \dot{m} } \bigr)
\ \bigl]. \label{norm}
\ee
(The $(\theta,\phi)$ parts of the stress-energy tensor are
determined by $G_{\theta\theta}$ and $G_{\phi\phi}=\sin^2\theta
 G_{\theta\theta}$, and are already diagonal
(since $G_{\theta\phi}\equiv 0$).
The corresponding eigenvectors are spacelike.)

Stress-energy tensors are classified by their
eigenvectors, and  for physical fields  this tensor must
be either Type I, for which there is one   timelike and
  three  spacelike eigenvectors, or Type II, for which
there are two null  and two spacelike eigenvectors
\cite{hawkell}.  All physical fields are of type I, except
for certain null fluid flows,  which are of type II.

We will focus on the type I tensors. Let $-\rho$ be the
eigenvalue corresponding to the timelike eigenvector,
and $\pi_i$ $(i=1,2,3)$ the eigenvalues corresponding to the
three spacelike eigenvectors. Then the  weak energy condition,
which requires that the energy density be non-negative,
is
\be
\rho \ge 0,  \ \ \ {\rm and} \ \ \ \pi_i \ge -\rho \ \ \
 {\rm for}\ \ \  i=1,2,3.
\label{weak}
\ee
The dominant energy condition, which requires that energy
flows are never spacelike, imposes in addition to (\ref{weak}),
the condition
\be
\pi_i \le \rho \ \ \   {\rm for}\ \ \  i=1,2,3\ .
\ee

{}From our chosen mass function (\ref{ourm}), we see that
$\dot{m}>0$, which requires that $\psi^\prime >0$ for
the square root in the eigenvalues (\ref{eval}) to be real.
We note from (\ref{eval}) and (\ref{norm}) that
 $ 4\dot{m} e^{-\psi}/ r f^2 \psi^\prime > 0$, (which is
always true),  implies that the
eigenvector $u^{(-)}$ is timelike as required.  Also,
$u^{(+)}$ is always spacelike.  Therefore, for the weak
energy condition we need a $\psi$ that satisfies
$\psi^\prime >0$. In addition, to preserve the
asymptotic flatness of the metric we require
\be
\lim_{r\rightarrow\infty} \psi(r,v=\infty)= constant.
\ee
Since we already have $\lambda_+ > \lambda_-$, for the
weak energy condition we require, in addition, a $\psi$
 such that
\be
\lambda_- \le 0,
\ee
and
\be
G_{\theta\theta} = (r + m) \psi^\prime
- 3 r m^\prime \psi^\prime
 + r^2 f (\psi^\prime)^2
- r^2 e^{-\psi} \dot{\psi}^\prime
 - r m^{\prime\prime}
+ r^2 f \psi^{\prime\prime} \ge \lambda_-.
\ee
For the dominant energy condition we also require
\be
 -\lambda_- \ge \lambda_+ \ \ \ {\rm and} \ \ \
-\lambda_- \ge G_{\theta\theta}\ ,
\ee
that is, all the pressures must be bounded between
$\rho$ and $-\rho$.

A $\psi$ that satisfies both the weak and dominant
energy conditions is of the form
\be
\psi(r,v) = -e^{- r f(v)} - g(v),
\label{psi}
\ee
where the functions $f$ and $g$ are everywhere positive
and have a lower bound  $>$ 2. This last condition is necessary
to enforce the dominant energy condition, which
was checked to be true numerically for $r \ge$ horizon
radius, and for subcritical and supercritical ranges of
$c$.

Our main result is a {\it general} method for obtaining a mass
formula for black holes via solutions to a transcendental
equation: given any explicit collapse solution, a black hole
mass formula  must arise in this way, (and it may or may not be
a pure power law).  As an explicit example of the procedure, we
have given a metric that describes a physically realistic
spherically symmetric collapse, and that exhibits a mass gap
at the onset of black hole formation.

  It would be of interest to see if there are other choices for
the mass function that describe a realistic collapse.
In particular, it would be of much interest to find
the mass functions that give gapless black hole mass formulas.

An important question \cite{ted} concerning critical behaviour
in our approach is what features of the mass function are
responsible for giving criticality. The question actually has
at least two parts: What features give the three subcritical,
critical, and supercritical regimes, and what features give
rise to a pure power law, and hence a critical exponent? The answer
to the first question is that
any mass function that grows with the parameter in approximately
the same way as ours is sufficient. We speculate that the
answer to the second question requires an approximately
step shaped mass function (like ours) such that its knee remains
{\it just touching} the $r/2$ curve (Fig. 2) for a wide range
of parameter  values $c$, (instead of crossing it, and thereby
giving an inner horizon also, as ours does).
In our approach the former choice of mass function, (that just
touches the $r/2$ curve), requires
`fine tuning' whereas the second choice appears to be more
generic.
  A deeper question is
what properties of the Einstein equations in this context
give rise to mass functions having these features.

A further point to note is that critical behaviour may
occur even for unrealistic matter: we simply leave
$\psi$ arbitrary, up to requiring  asymptotic flatness of
the metric, and not worry about imposing the energy
conditions.

It seems possible to use this approach for axially-symmetric
metrics as well; one could for example use the mass function
given here, and still have two remaining metric functions at
hand for satisfying the dominant energy condition.

\bigskip
We would like to thank Werner Israel, Ted Jacobson, Karel Kuchar,
Don Page, Richard Price, and Lee Smolin for very helpful comments.
The work of V. H. and E. M. was supported by the
Natural Science and Engineering Research Council of
Canada, and that of D. N. partly by DGAPA, National
A. University of M\'exico.

\vfill
\eject

\noindent FIGURE CAPTIONS
\medskip

\noindent Figure 1.  $M_{BH}$ vs. ($c-c_*)$ curve whose best fit
is Eqn. (12).

\medskip
\noindent Figure 2. The critical mass function ($c=.46573$) together
with a typical subcritical case (c=.3). The other curve is the
function $r/2$ which touches the critical mass function.

\end{document}